\documentclass[english,prl,twocolumn,superscriptaddress,showpacs]{revtex4}

\usepackage{graphicx}
\usepackage{dcolumn}
\usepackage{bm}

\begin{document}

\title{Murphy et al. Reply to the Comment by Kopeikin on \\``Gravitomagnetic Influence on Gyroscopes and on the Lunar Orbit''} 

\author{T.\,W.~Murphy,~Jr.}
\email[]{tmurphy@physics.ucsd.edu}
\affiliation{University of California, San Diego, 9500 Gilman Drive,
La Jolla, CA 92093-0424}
\author{K.~Nordtvedt}
\affiliation{Northwest Analysis, 118 Sourdough Ridge Road, Bozeman, MT 59715}
\author{S.\,G.~Turyshev}
\affiliation{Jet Propulsion Laboratory, California Institute of Technology,
4800 Oak Grove Drive, Pasadena, CA 91109}

\date{\today}

\maketitle

In \cite{gravmag}, we point out that a gravitomagnetic term in the equation
of motion used to dynamically determine the precise shape of the lunar
orbit is also responsible for the ``frame-dragging" precession of a
gyroscope near a massive rotating body.  In the former case, the
gravitomagnetic interaction between the moving masses of Earth and
Moon---as evaluated in the solar system barycenter (SSB) frame---leads to
orbital amplitude contributions at the six meter level.  Part of the
gravitomagnetic interaction plays a role in producing the necessary Lorentz
contraction of the orbit in this frame.  In the case of a gyroscope, the
same interaction between mass elements moving within the macroscopic bodies
produces the gyroscope precession.  The physics is the same for both, the
difference being in the distribution of mass currents.

The SSB frame is chosen for lunar laser ranging (LLR) analysis for a
variety of practical reasons \cite{Williams-etal-2005}---not least of which
that it is the most convenient asymptotically inertial reference frame for
solar system dynamical analyses.  The lunar and planetary orbits and the
lunar rotation are determined by a simultaneous numerical integration of
the post-Newtonian differential equation of motion and evaluation of light
propagation times between the moving Earth and Moon
\cite{WilliamsStandish2006}.

The six-meter gravitomagnetic influences on the lunar orbit in the SSB
frame appear as $\cos D$ and $\cos 2D$ signatures, where $D$ is the synodic
phase.  The post-Newtonian model, as implemented in the way described
above, fits decades of LLR data in these modes to 4~mm and 8~mm accuracy,
respectively.  Therefore, any \emph{isolated} modification of the
gravitomagnetic term is limited to $\approx 0.1$\% the strength prescribed
by general relativity \cite{gravmag}.

The gravitomagnetic term in the equation of motion is just one of several
velocity-dependent contributions to the whole \cite{jgw96}.  It is
physically unrealistic to adjust the strength of a single interaction term
without simultaneously examining changes to other terms in the velocity
transformation package.  Self-consistent transformations of the
velocity-dependent terms from one frame to another in a metric
framework have been worked out \cite{ppn}, and strongly constrained by
experiment at well below the 0.1\% level relevant to this discussion
\cite{alpha1,alpha2}.

It is clear that the choice of reference frame affects the lunar orbit
shape needed to fit the LLR ranging data---and specifically the
gravitomagnetic interaction's contribution to that frame-dependent orbit
\cite{kopeikin}.  Also clear is that current successful LLR analysis
performed in the SSB frame requires inclusion of general relativity's
prescribed gravitomagnetism.  Therefore, this interaction cannot be
arbitrarily adjusted---alone or together with other aspects of post-Newtonian
gravity---without considering the impact of such adjustments on LLR as well
as on the variety of other relevant observations such as ranging to Mars and
Mercury, binary pulsar pulse arrival times, etc.

\end{document}